# Nitrogen-doped Nanoporous Carbon Membranes Functionalized with Co/CoP Janus-type nanocrystals as Hydrogen Evolution Electrode in Both Acid and Alkaline Environment

*Hong Wang [†,‡], Shixiong Min[#,‡], Qiang Wang[§], Debao Li[§], Gilberto Casillas[⊥], Chun Ma[†], Yangyang Li[†], Zhixiong Li[†], Lain-Jong Li[†], Jiayin Yuan[∥,#,\*], Markus Antonietti[∥], Tom Wu[†]\**

[†]Physical Science and Engineering Division, King Abdullah University of Science & Technology (KAUST), Thuwal, 23955-6900, Saudi Arabia

[#]School of Chemistry and Chemical Engineering, Beifang University of Nationalities, Yinchuan, Ningxia, China

[§]State Key Laboratory of Coal Conversion, Institute of Coal Chemistry, the Chinese Academy of Sciences, Taiyuan 030001, China

[⊥]UOW Electron Microscopy Centre, University of Wollongong, Wollongong, New South Wales 2500, Australia

[∥]Department of colloidal chemistry, Max Planck Institute of Colloids and Interfaces, 14476 Potsdam, Germany

[#]Department of Chemistry and Biomolecular Science & Center for Advanced Materials Processing, Clarkson University, 13699, USA

ABSTRACT. Self-supported electrocatalysts being generated and employed directly as electrode for energy conversion has been intensively pursued in the fields of materials chemistry and energy. Herein, we report a synthetic strategy to prepare freestanding hierarchically structured, nitrogen-doped nanoporous graphitic carbon membranes functionalized with Janus-type Co/CoP nanocrystals (termed as HNDCM-Co/CoP), which were successfully applied as a highly-efficient, binder-free electrode in hydrogen evolution reaction (HER). Benefited from multiple structural merits, such as high degree of graphitization, three-dimensionally interconnected micro-/meso-/macropores, uniform nitrogen-doping, well-dispersed Co/CoP nanocrystals as well as the confinement effect of the thin carbon layer on the nanocrystals, HNDCM-Co/CoP exhibited superior electrocatalytic activity and long-term operation stability for HER under both acid and alkaline conditions. As a proof-of-concept of practical usage, a macroscopic piece of HNDCM-Co/CoP of 5.6 cm x 4 cm x 60 μm in size was prepared in our laboratory. Driven by a solar cell, electroreduction of water in alkaline condition (pH 14) was performed, and $H_2$ has been produced at a rate of 16 ml/min, demonstrating its potential as real-life energy conversion systems.

Hydrogen is a clean, renewable energy carrier and has been actively pursued as an alternative to fossil fuels[1-3]. Sustainable production of hydrogen from water splitting is an appealing solution yet requires highly efficient, long-term stable electrocatalyst[4, 5]. As a benchmark electrocatalyst, platinum is sufficiently active in hydrogen evolution reaction (HER)[6], but suffers from high cost and scarcity. As technological breakthroughs, a variety of non-noble metal catalysts, *e.g.* $Ni_2P$[7], $Mo_2C/MoB$[8], $CoO_x$[9], CoPS[10], $MoS_2$ and its hybrids[11, 12, 13], CoP/CNTs[14], alloys[15], and $CoSe_2$ and its carbon hybrids[16, 17], just to name a few, have been explored for HER. Among these systems, carbon-based hybrids were found to outperform pristine metal ones due to the following factors that are still under debate: i) chemical and electrical coupling effects (charge transfer, heterojunction, *etc.*) between metallic electrocatalyst and conductive carbon[18, 19]; ii) better accessible active centers of metal species endowed by the conductive porous carbon scaffold[14, 20]. The choice of carbon nanostructures is in this regard critical for design and engineering of HER electrocatalyst, and usually powdrous carbon nanostructures such as reduced graphene oxide[21-25], carbon nanotubes[14, 26], N-doped carbons[9, 27], are mostly applied in the field. In spite of their favorable efficiency in $H_2$ generation, device-wise the powderous electrocatalysts have to be engineered into electrodes of defined shape usually *via* the usage of a polymer binder, such as Nafion or polyvinylidene fluoride. The use of binder is a mature processing practice in industry for decades, but their presence leads to multiple undesirable effects, including reduced cell conductivity, weak polymer/carbon/metal heterojunction, and blocking of the active centers or restricted diffusion that leads to reduced catalytic activity. Additionally, the side reactions of polymer binders during the electrochemical process are a rising concern[28].

The problem of electrode integrity is solved by applying a macroscopic carbon membrane supported electrocatalyst as a binder-free electrodes for HER, in which the synthesis of high-quality porous carbon membranes is the key feature. It is well-recognized that the carbon membrane design should entail a hierarchical pore architecture over broad length scales from micro- to meso- to macropores. The macropores afford rapid mass transport for the fluid and gas phase, and simultaneously the micro-/mesopores intrinsically contribute to a large specific surface area that provides enhanced reaction capacity for the heterogeneous reactions. In this context, the electrocatalyst reaches an optimized balance between activity and diffusion kinetics[29]. Equally important, heteroatoms, such as nitrogen, when covalently bound in the carbon network, are able to tailor and improve material properties of carbons such as conductivity, chemical inertness and basicity, which are beneficial to their catalytic function in HER[22, 30, 31]. These desirable features raise structural and synthetic complexity and challenge the design of carbon membrane-derived electrocatalyst for HER.

So far, most HER electrocatalysts are known to work efficiently under strongly acidic condition, simply because the large amount of $H^+$ facilitates the HER process for thermodynamic and kinetic reasons[32]. Alkaline conditions are more challenging to produce $H_2$ by electrolysis, as the system inherently suffers from high overpotentials and instability[33]. Nevertheless, generating hydrogen by electrolysis in alkaline media has been utilized for decades in industry[34, 35], as it offers high purity hydrogen and more importantly, the alkaline media simplifies the oxygen production side and cause less stability problems associated with the catalyst. Therefore, design of active, durable and low-cost electrocatalysts that can work well in alkaline media is highly desirable from the practical application viewpoint.

Herein, we report the synthesis of freestanding, hierarchically porous, nitrogen-doped graphitic carbon membranes loaded with Janus-type Co/CoP nanocrystals (termed HNDCM-Co/CoP), which demonstrate high activity towards HER in both acid and alkaline range. Electrochemical measurements showed a low overpotential of 135 mV and 138 mV at 10 mA cm$^{-2}$ in acid and alkaline conditions, respectively, and long-term durability over 20 h of HER operation in both acid and alkaline media using HNDCM-Co/CoP catalyst.

The synthetic route towards the targeted HNDCM-Co/CoP for HER electrocatalysis is displayed in **Figure 1**, in which a nanoporous polymer membrane (termed as NPPM) built up via interpolyelectrolyte complexation was used as a sacrificial soft template (**Figure 1a, 1c and Figure S1**) (see experimental section). A key feature of polyelectrolyte complexes is their capability to bind and immobilize metal ions, salts, and charged nanoparticles[36], which remarkably simplifies functionalization of the NPPM with metal nanoparticles. Here Co salt was chosen because, besides its relatively high abundance in nature, it has been theoretically demonstrated that some Co surfaces have a low energy barrier for H adsorption, and thus Co-based electrocatalysts hold a great promise in HER[37]. $Co^{2+}$ ions were immobilized in the NPM by refluxing the as-prepared NPM in an aqueous solution of cobalt acetate for 24 h. After rinsing with deionized water and drying, the NPM-cobalt acetate precursor was pyrolyzed at 1000 °C under $N_2$ flow to form Co nanoparticles functionalized carbon membranes (termed HNDCM-Co). The final HNDCM-Co/CoP (**Figure 1b, 1d**) with a shiny black color was obtained by phosphatization of HNDCM-Co using $NaH_2PO_4$ at 350 °C for 3 h under $N_2$ flow.

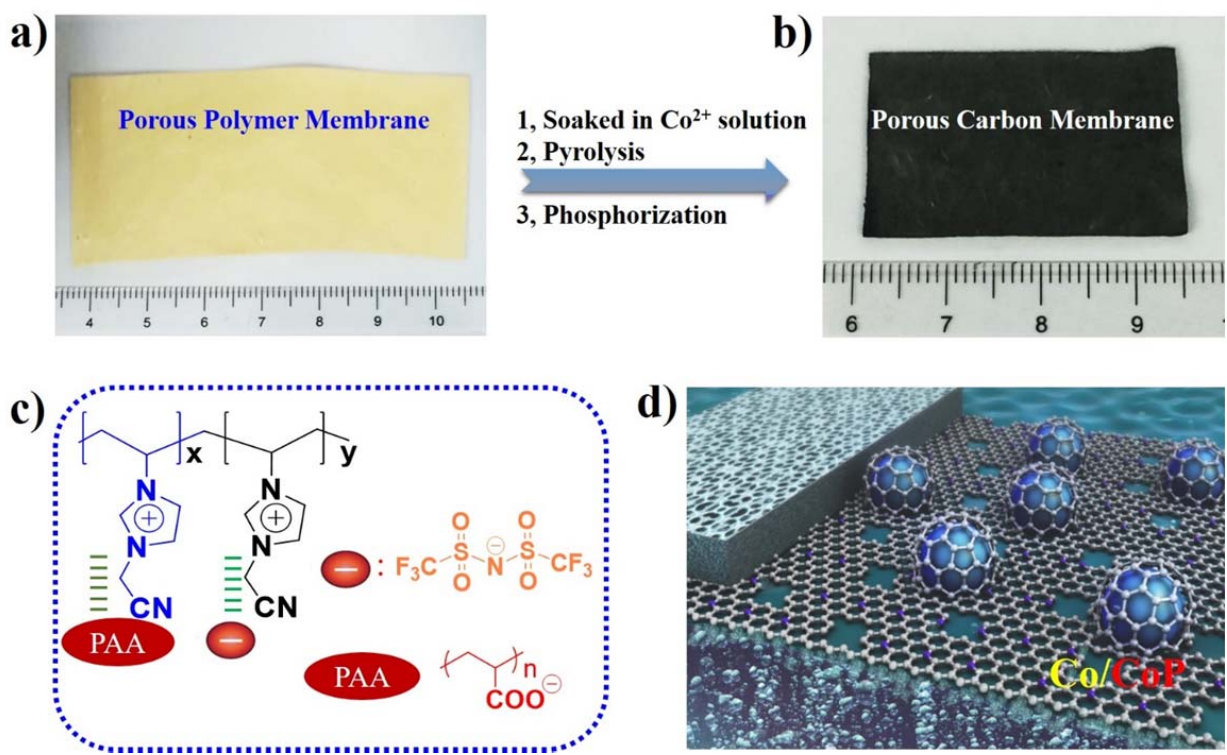

**Figure 1**, Synthetic procedure of hierarchically porous, nitrogen-doped graphitic carbon membranes loaded with Janus-type Co/CoP nanocrystals (HNDCM-Co/CoP). (a, b) Photographs of a freestanding NPPM of 7 x 3 $cm^2$, and a corresponding freestanding HNDCM-Co/CoP of 3.4

x 2 cm$^2$ in size. (c, d) the schematic presentation of the NPPM structure and the HNDCM-Co/CoP.

**Figure 2a** shows the SEM image of a cross-section of the HNDCM-Co/CoP hybrid membrane, in which the typical gradient of macropore size is clearly observable. The average macropore size in zone I, zone II, and zone III are 1.6 μm, 850 nm and 550 nm, respectively, i.e. the pore size gradually decreases from top to bottom. The three dimensionally interconnected cellular architecture can be clearly recognized in the enlarged SEM image in **Figure 2b**. On the pore walls, plenty of bright dots, i.e. the metal nanocrystals, are observed uniformly dispersed (denoted by red arrows, **Figure 2c**). Transmission electron microscopy (TEM) image (**Figure S2**) discloses the porous structure of the HNDCM-Co/CoP and confirmed again the uniform distribution of Co/CoP nanocrystals of 10~40 nm in size throughout the entire membrane. Energy-filtered transmission electron microscopy mappings for C, N, Co and P (**Figure 2d**) indicate a uniform distribution of N in the carbon matrix, which is expected due to in situ molecular doping of HNDCM with N[38]. It should be noted that P only exists in the region of Co, indicating that low temperature (350 °C) phosphatization of HNDCM-Co using NaH$_2$PO$_4$ could not lead to P-doped carbon. Surprisingly, after merging of elemental mappings of Co and P, it was found that majority of nanocrystals were composed of Co/CoP Janus-type nanostructures (**Figure 2e**). The phases of Co and CoP were further confirmed by using atomic High-angle annular dark-field scanning transmission electron microscopy (HAADF-STEM, **Figure 2f, 2g**). In order to understand the formation mechanism of Janus-type structure of Co/CoP, a single Co/CoP Janus-type nanocrystal was analyzed by HRTEM (**Figure S3**). It was found that the region of Co is protected by a few nm thin graphitic layer, while the CoP region is exposed to surrounding. Therefore, it is reasonable to conceive that phosphatization reaction firstly occurred on the exposed surfaces of Co nanoparticles and the graphitic carbon protected Co is difficult to be phosphatized, thus forming such a Janus-type structure of Co/CoP. According to previous reports[39], the close contact between metal (Co in our case) and N-doped carbon is effective in producing a rectifying effect (**Figure S4**), which can polarize/activate the interface and improve the final catalytic performance.

The phase of HNDCM-Co/CoP was further confirmed by X-ray diffraction (XRD) and X-ray photoelectron spectroscopy (XPS), as provided in **Figure S5**. The total Co content was 42.8 mg/g, as detected by inductively coupled plasma-atomic emission spectra. The relative content of Co and CoP in Janus-type Co/CoP nanocrystals can be calculated from the XPS spectra (**Figure S5c**) as 13.3 mg/g and 29.5 mg/g, respectively. Furthermore, a HRTEM image (**Figure S6**) shows that a disordered graphitic character is maintained throughout the entire porous membrane. The well-developed graphitic carbon membrane endows the HNDCM-Co/CoP with favorably high conductivity of 51 S cm$^{-2}$ at 25 $^{\circ}$C (**Figure S7**). The high conductivity of HNCM-Co/CoP favors fast charge transport, a mandatory requirement for efficient electrocatalysis. The content of N in NHDCM-Co/CoP is 5.4 wt%, as determined by elemental analysis. All of these results point out that Janus-type Co/CoP nanoalloy crystals were formed and embedded in the N-doped porous carbon membrane via the low-temperature phosphatization of HNDCM-Co. The functionalization of transition-metal nanocomposite-based electrocatalysts, improving the compatibility of these electrocatalysts with electrolytes, is currently one of the most important research topics in the field of electrocatalysis[39]. Mott–Schottky catalysts, made of metal–semiconductor heterojunctions, have been recently applied to dehydrogenation, artificial photosynthesis, and eletrocatalysis[40, 41]. It is a consensus that the Mott–Schottky effect at the metal–semiconductor interfaces of Janus-type nanoparticles can dramatically promote the final performance and stability owing to possible synergetic effects and enhanced electron transfer efficiency at the interface between different components (**Figure S8**).

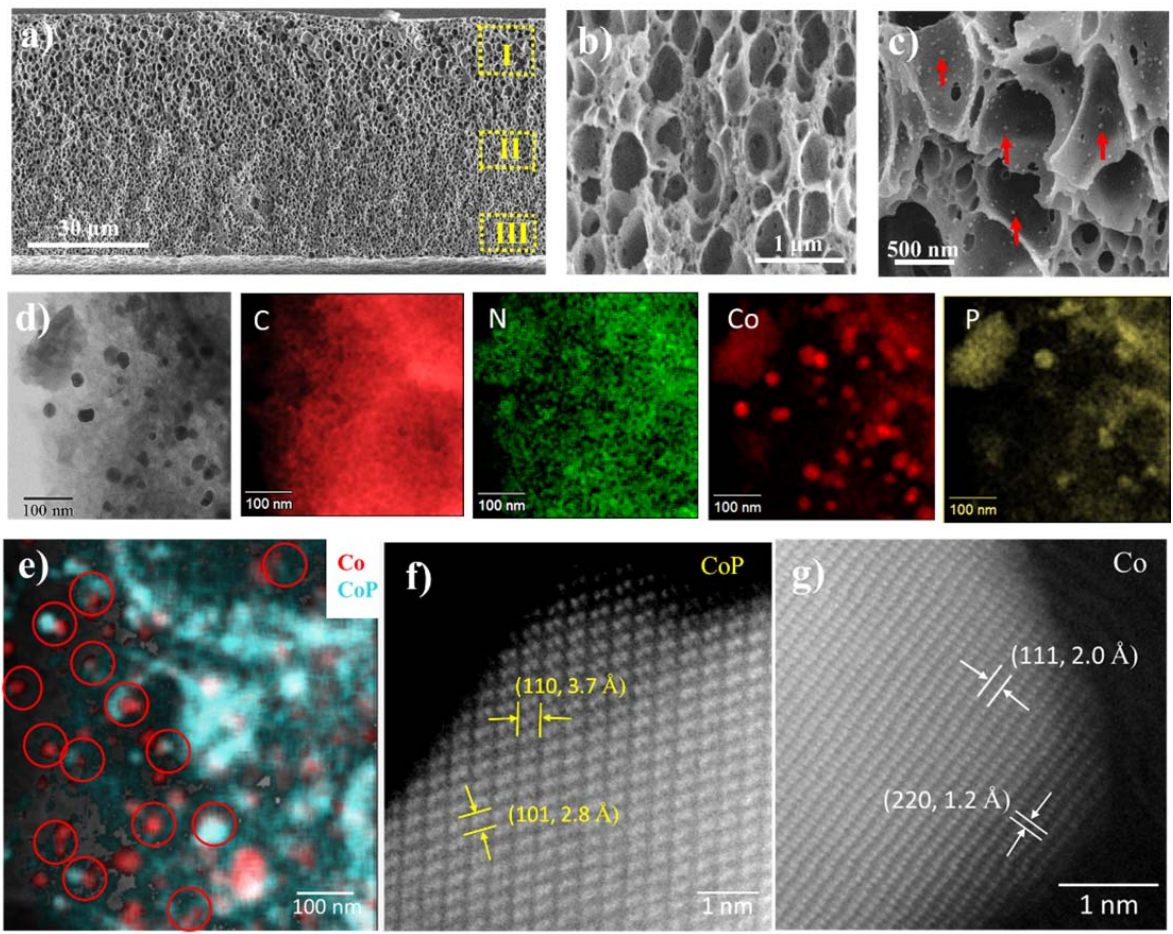

**Figure 2.** (a-c), Overview and close view of the cross-sectional SEM images of HNDCM-Co/CoP. (d) Energy-filtered transmission electron microscopy mappings for C, N, Co and P. (e) Merged elemental mappings of Co and P. Red circles indicate Janus-type nanostructures of Co/CoP. Also shown are the HAADF-STEM images of (f) CoP and (g) Co phases in Janus-type nanocrystals.

In addition, the Brunauer-Emmett-Teller (BET) specific surface area of HNDCM-Co/CoP was determined by nitrogen gas sorption to be 589 $m^2/g$ (**Figure 3a**). The sharp increase of BET at low pressures (P/P$_0$ < 0.05) is due to the nitrogen filling in micropores below 2 nm, which is quantified by the density functional theory (DFT) pore size distribution curves (**Figure 3b**) derived from the N$_2$ adsorption branches. The obvious hysteresis above P/P$_0$ ~ 0.5 is indicative of the existence of mesopores. The pore volumes of the micropores and mesopores were 0.07 and 0.58 cm$^3$ g$^{-1}$, respectively. It is clear that the porous carbon membrane features not only a gradient in the macropore size distribution along its cross-section, but is also simultaneously rich in micro- and mesopores. As previously mentioned, the large macropores provide transport highways while the micro- and mesopores offer the necessary large specific surface area bearing active sites for the heterogeneous reactions. It should be noted that this pore texture in the porous carbon membrane is obtained in a single carbonization step without any post-synthesis activation treatment. Owing to the high conductivity, satisfactory BET surface area, hierarchical pore architecture as well as evenly dispersed Co/CoP nanocrystals, HNDCM-Co/CoP is well suited for many electrochemical processes, and it will be exemplified in the model case of water splitting as discussed beneath.

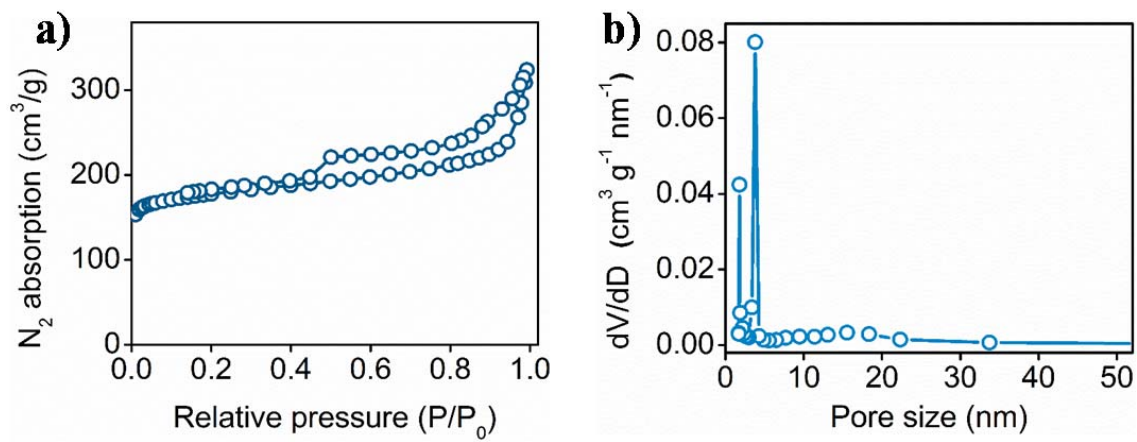

**Figure 3**. (a) N$_2$ absorption-desorption isotherms and (b) the corresponding pore size distribution of HNDCM-Co/CoP.

The HER activity of HNDCM-Co/CoP was evaluated by a standard three-electrode electrochemical cell in both acid and alkaline conditions and was compared with the metal-free carbon membrane HNDCM, and only Co nanoparticle-functionalized carbon membrane HNDCM-Co. It is noted that the size and thickness of the three electrocatalysts as well as the membranes textures are almost equal. All HER data has been corrected based on impedance spectroscopy, as shown in **Figure 4**. At 10 mA/cm$^2$, HNDCM and HNDCM-Co exhibited overpotentials of 823 and 247 mV, respectively. Under the exact same condition, HNDCM-Co/CoP showed the lowest overpotential of 138 mV, which is in fact one of the best non-noble-metal electrocatalysts reported so far for HER (**Figure 4a**) (**Table S1**). In a basic environment, i.e., in 1 M KOH (pH 14), the LSV curves (**Figure 4b**) present overpotentials of 723 and 216 mV, respectively, for HNDCM and HNDCM-Co at 10 mA/cm$^2$, which are slightly lower than that in acid condition. It indicates that alkaline conditions are more favorable for N-doped carbon based samples than acidic conditions, also for a HER operation. HNDCM-Co/CoP requires an overpotential as low as 135 mV, close to that in acidic condition. It should be noted that the 'noise' in LSV curves for HER in both of acid and alkaline conditions were generated by perturbations in our membrane catalyst due to the release of large amounts of H$_2$ bubbles that were produced at higher overpotentials. Thus far, there are only a few electrocatalysts active in both acid and alkaline conditions, due to the incompatibility of the activity of the same electrocatalyst operating in the same pH region[42, 43]. The activities for HER in alkaline are usually about 2~3 orders of magnitude lower than those in acidic medium[44]. In our case, the excellent HER activity of HNDCM-Co/CoP in alkaline condition can be potentially attributed to the multiple heterojunctions, i.e. the support interaction with the N-doped carbon as well as the bi-phasic character of Co and CoP in the Janus nanocrystals.

Tafel plots were used to elucidate the electron-transfer kinetics. The linear portions of the Tafel plots (**Figure 4c**) are fitted to the Tafel equation: η=b logj + a, where j is the current density and b is the Tafel slope. The Tafel slopes of HNDCM-Co/CoP are determined to be approximately 64 and 66 mV dec$^{-1}$ in 0.5 M $H_2SO_4$ and 1M KOH, respectively. This result suggests that the HER over HNDCM-Co/CoP followed a Volmer-Heyrovsky mechanism in both acid and alkaline conditions, and the electrochemical desorption step is rate-limiting[14,45]

Additionally, stability of electrocatalysts in practical operation conditions is a key parameter. We investigated the long-term electro-chemical stability of HNDCM-Co/CoP for HER in both acid and alkaline conditions (**Figure 4d**), and no decay was detected during continuous operation of 20 h. Meanwhile, the cyclic voltammetry (CV) durability tests of the HNDCM-Co/CoP electrodes for HER in both of acid and alkaline conditions were carried out, as shown in **Figure S9**. Obviously, the HNDCM-Co/CoP electrodes exhibited negligible loss of activity after 1000 CV sweeps. Herein, N-doping leads to the basicity of the HNDCM-Co/CoP, which could improve the electrochemical stability and resistance against oxidation by modifying the electronic band structure of the graphitic carbons[46], integrated porous membrane structure (binder-free), and probably the confinement effect of the thin carbon layer on the nanocrystals endow HNDCM-Co/CoP with excellent stability.

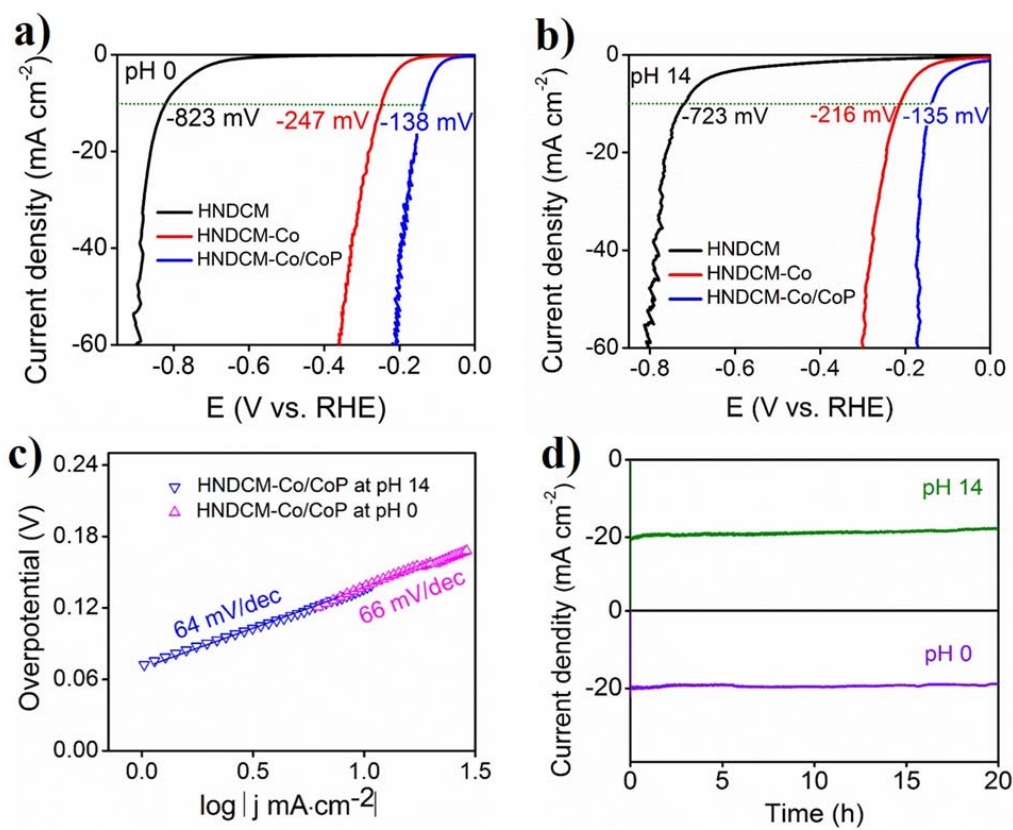

**Figure 4**. HER performances of HNDCM, HNDCM-Co and HNDCM-Co/CoP in (a) 0.5 M H$_2$SO$_4$ and (b) 1 M KOH. (c) Tafel plots of HNDCM-Co/CoP in acid and alkaline conditions. (d) Stability of HNDCM-Co/CoP in 0.5 M H$_2$SO$_4$ (pH 0) and 1 M KOH (pH 14).

As an important feature, our synthetic route towards freestanding membrane-type HER electrode can be easily scaled up. As a proof-of-concept demonstration for the solar-driven electrolysis of water, a large piece of HNDCM-Co/CoP of 5.6 x 4 cm$^2$ in size and 60 μm in thickness (**Figure 5**) was prepared and driven by a solar panel (up to 20 V) for HER in 1 M KOH, at an output voltage of non-regulated, fluctuating 20 V. Such extreme and fluctuating conditions usually destroy any ordinary electrocatalytic material in short times. It is also noted that this membrane was produced in a laboratory carbonization oven, but essentially any larger size can be made, given a corresponding carbonization technology. **Figure 5** illustrates the HER operation within 10 min, and 160 mL H$_2$ was collected, indicating the practicality of HNDCM-Co/CoP in simple, decentral H$_2$ production in an environment-friendly manner.

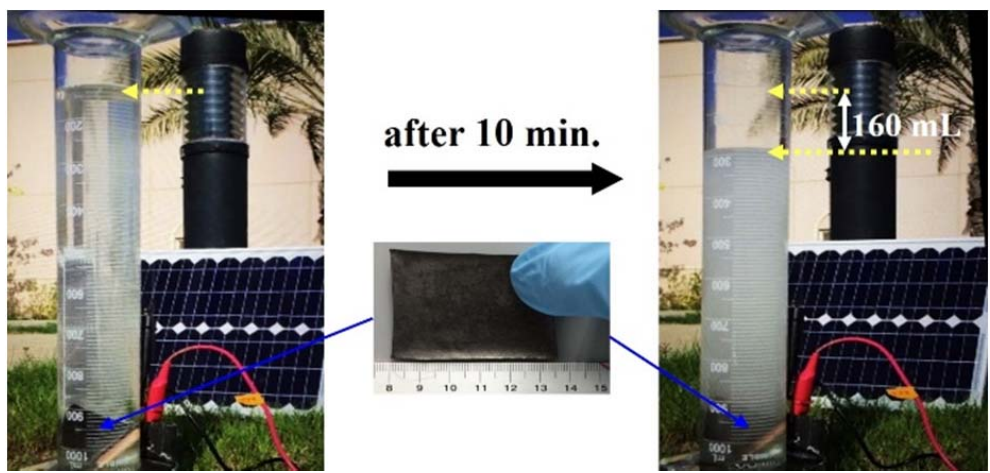

**Figure 5.** Illustration of HER driven by a solar panel in 1 M KOH with a piece of hybrid membrane HNDCM-Co/CoP with a size of 5.6 x 4 cm$^2$ and a thickness of 60 μm as working electrode and the anode is a graphite electrode. In 10 min, 160 mL of H$_2$ was released from HNDCM-Co/CoP membrane.

**CONCLUSIONS**

In summary, we presented a general and scalable method to prepare N-doped, hierarchically structured porous graphitic carbon membranes with self-supported Janus-type Co/CoP nanocrystals, HNDCM-Co/CoP serves as a highly active and robust earth-abundant electrode for hydrogen evolution reaction in both acid and alkaline conditions. Ease of large-scale preparation in combination with excellent electroactivity and remarkable long-term operation stability make HNDCM-Co/CoP promising for industrial-scale HER application. Another unique feature of this genre of polyelectrolyte membrane is the ability to incorporate any type of metal ion and nanoparticle into the structure. Thereafter, metal nanoparticle functionalized HNCM membranes can be readily prepared by carbonization of a metal/polyelectrolyte membrane. We envision different metal functionalized N-doped carbon membrane will provide myriad opportunities to develop highly efficient noble-metal-free electrocatalysts for power-to-fuel conversions, as discussed in current sustainable energy system solutions.

EXPERIMENTAL SECTION

**Materials and Reagents.** 1-Vinylimidazole (Aldrich 99%), 2,2'-azobis(2-methylpropionitrile) (AIBN, 98%), bromoace-tonitrile (Aldrich 97%), and bis(trifluoromethanesulfonyl)imide lithium salt (LiTFSI, Aldrich 99%) were used as received without further purifica-tions. Solvents of analytic grade were used as received. Poly(acrylic acid) (PAA) MW: 100,000 g/mol, 35 wt% in water, was obtained from Sigma Aldrich and used in a powder form after freeze-drying. The synthetic procedure of NPM was according to our previous method[47].

**Electrochemical Characterizations.** The electrochemical measurements were performed with an electrochemical im-pedance spectroscopy (EIS) capable channel in a Biologic VMP3 potentiostat. A graphite rod and an Ag/AgCl (in saturated KCl solution) electrode were used as the counter and reference electrodes, respectively. All the applied poten-tials are converted to reversible hydrogen electrode (RHE) potentials scale by using equation: E (vs. RHE) = E (vs. Ag/AgCl) + 0.217 V + 0.0591 V*pH, after IR correction. Potentiostatic EIS was used to determine the uncompensated solution resistance ($R_s$). The HER activity of HNDCM-Co/CoP was evaluated by measuring polarization curves with linear sweep voltammetry (LSV) technique at a scan rate of 1 mV/s in 1.0 M KOH (pH 14) and 0.5 M $H_2SO_4$ (pH 0) solutions. The stability tests for the HNDCM-Co/CoP were performed using chronoamperometry at a constant applied overpotential.

**Characterizations.** X-ray diffraction (XRD) patterns were obtained using a Rigaku powder X-ray diffractometer with Cu K$\alpha$ ($\lambda$ = 1.5418 Å) radiation, 2$\theta$ angel was recorded from 20 to 80 degree. X-ray photoelectron spectroscopy (XPS) data were collected on an Axis Ultra instrument (Kratos Analytical) under ultrahigh vacuum (<10$^{-8}$ Torr) using a monochromatic Al Kα X-ray source. The adventitious carbon peak was calibrated at 285 eV and used as an internal standard to compensate for any charging effects. A field emission scanning electron microscope (FESEM, FEI Quanta 600FEG) was used to acquire SEM images. Transmission electron microscope (TEM) and high resolution TEM (HRTEM) images, selected-area electron diffraction (SAED) patterns, and the HAADF-STEM-EDS data were taken on a JEOL JEM-2100F transmission electron microscopy operated at 200 kV. Nitrogen sorption isotherms were measured at -196 °C using a Micromeritics ASAP 2020M and 3020M system. The samples were degassed for 6 h at 200 °C before the measurements. STEM images were acquired in a probe-corrected JEOL ARM200F operated at 80 kV equipped with a cold field emission gun and a high resolution pole-piece.

ASSOCIATED CONTENT

**Supporting Information**. SEM image of NPM; TEM image of HNDCM/Co/CoP; HRTEM image of Co/CoP Janus nanocrystal; HRTEM image of graphtic N-doped carbon; Temperature dependence of conductivity of HNDCM-Co/CoP; XRD patterns of HNDCM-Co and HNDCM-Co/CoP; XPS spectrum of HNDCM-Co/CoP; Table S1 for comparison the performance of different electrocatalysts (PDF). This material is available free of charge via the Internet at http://pubs.acs.org.

**Corresponding Authors**

jyuan@clarkson.edu;

tao.wu@kaust.edu.sa

**Author Contributions**

‡These authors contributed equally.

**Notes**

The authors declare no competing financial interest

ACKNOWLEDGMENT

H. W. and T. W. thank the King Abdullah University of Sci-ence and Technology (KAUST) for financial support. S. M acknowledges the financial support from the National Natural Science Foundation of China (21463001). J. Y. is grateful for financial support from the Max Planck society, Germany, and the ERC (European Research Council) Starting Grant (project number 639720-NAPOLI).

# Supporting Information

**Nitrogen-doped Nanoporous Carbon Membranes Functionalized with Co/CoP Janus-type nanocrystals as Hydrogen Evolution Elec-trode in Both Acid and Alkaline Environment**

*Hong Wang [†‡], Shixiong Min[#‡], Qiang Wang[§], Debao Li[§], Gilberto Casillas[⊥], Chun Ma[†], Yangyang Li[†], Zhixiong Li[†], Lain-Jong Li[†], Jiayin Yuan[∥#\*], Markus Antonietti[∥], Tom Wu[†]\**

[†]Physical Science and Engineering Division, King Abdullah University of Science & Technology (KAUST), Thuwal, 23955-6900, Saudi Arabia

[#] School of Chemistry and Chemical Engineering, Beifang University of Nationalities, Yinchuan, Ningxia, China

[§]State Key Laboratory of Coal Conversion, Institute of Coal Chemistry, the Chinese Academy of Sciences, Taiyuan 030001, China

[⊥]UOW Electron Microscopy Centre, University of Wollongong, Wollongong, New South Wales 2500, Australia

[∥]Department of colloidal chemistry, Max Planck Institute of Colloids and Interfaces, 14476 Potsdam, Germany

[#]Department of Chemistry and Biomolecular Science & Center for Advanced Materials Processing, Clarkson University, 13699, USA

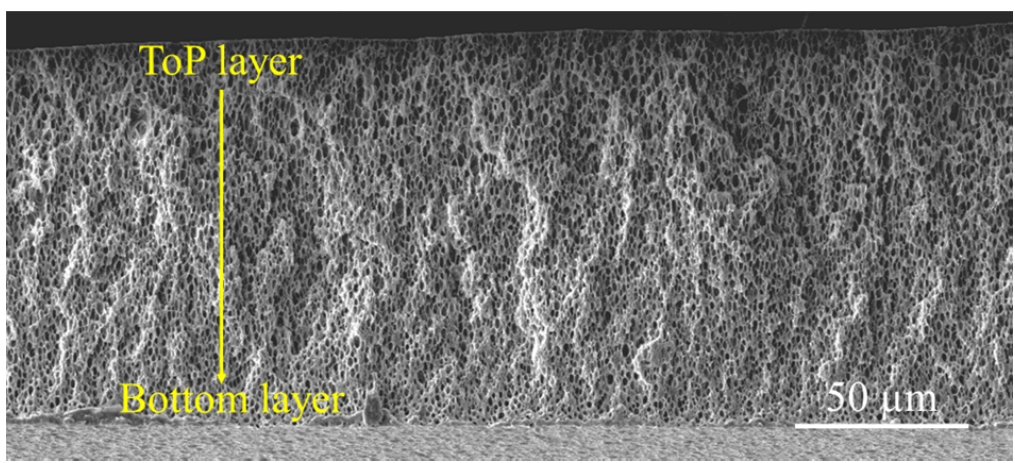

**Figure S1.** SEM image of a polymer membrane, which is used as sacrifice for synthesis of the gradient porous carbon membrane. The macropore gradually decreased from the top layer to the bottom layer.

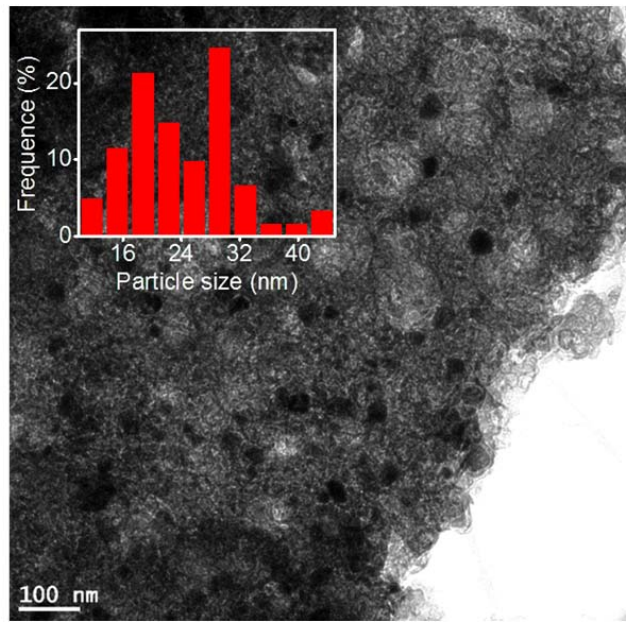

**Figure S2.** TEM image of HNDCM-Co/CoP. Inset is the number-averaged size distribution histogram of Co/CoP Janus nanoparticles

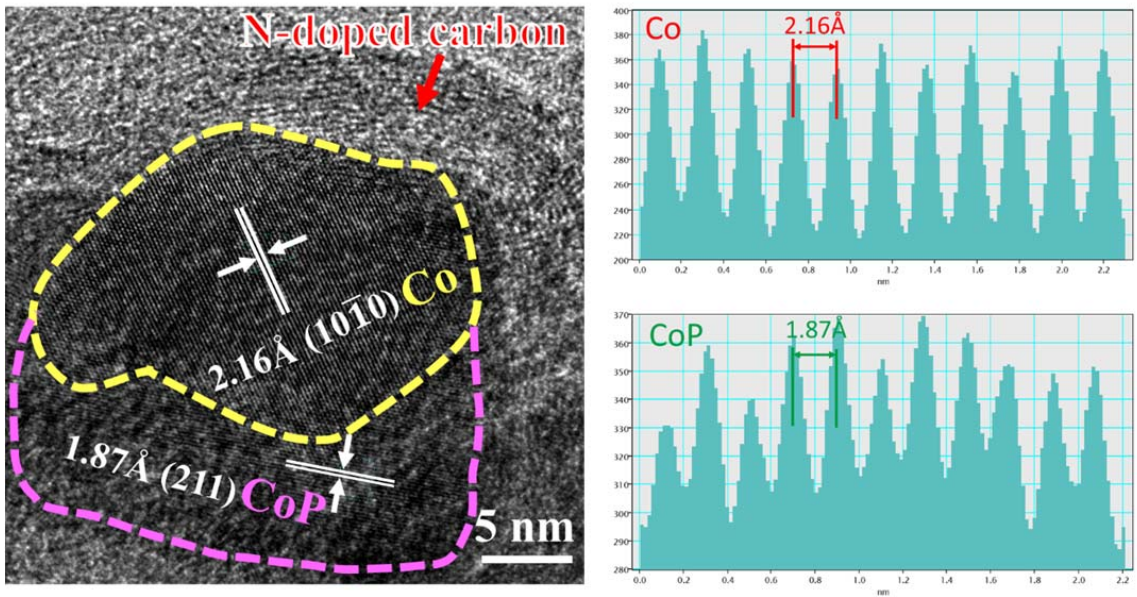

**Figure S3.** HRTEM image of a Janus Co/CoP nanoparticle. The yellow line-dotted area is composed of metallic Co with a lattice d-spacing of 2.16 Å, corresponding to the {10$\bar{1}$0} plane of fcp Co, while the adjacent pink line-dotted area stems from CoP with a lattice d-spacing of 1.87 Å from the (211) plane of CoP, *i.e.* the nanocrystal is of Janus-type. It can be clearly seen that the region of Co is protected by a few nm thin graphitic carbon, and the thickness of the graphitic carbon is about 8 nm.

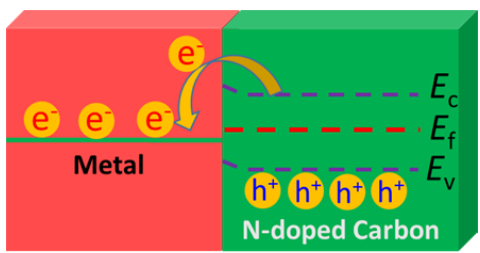

**Figure S4.** The rectifying effect arising from the close contact of metal and N-doped carbon.

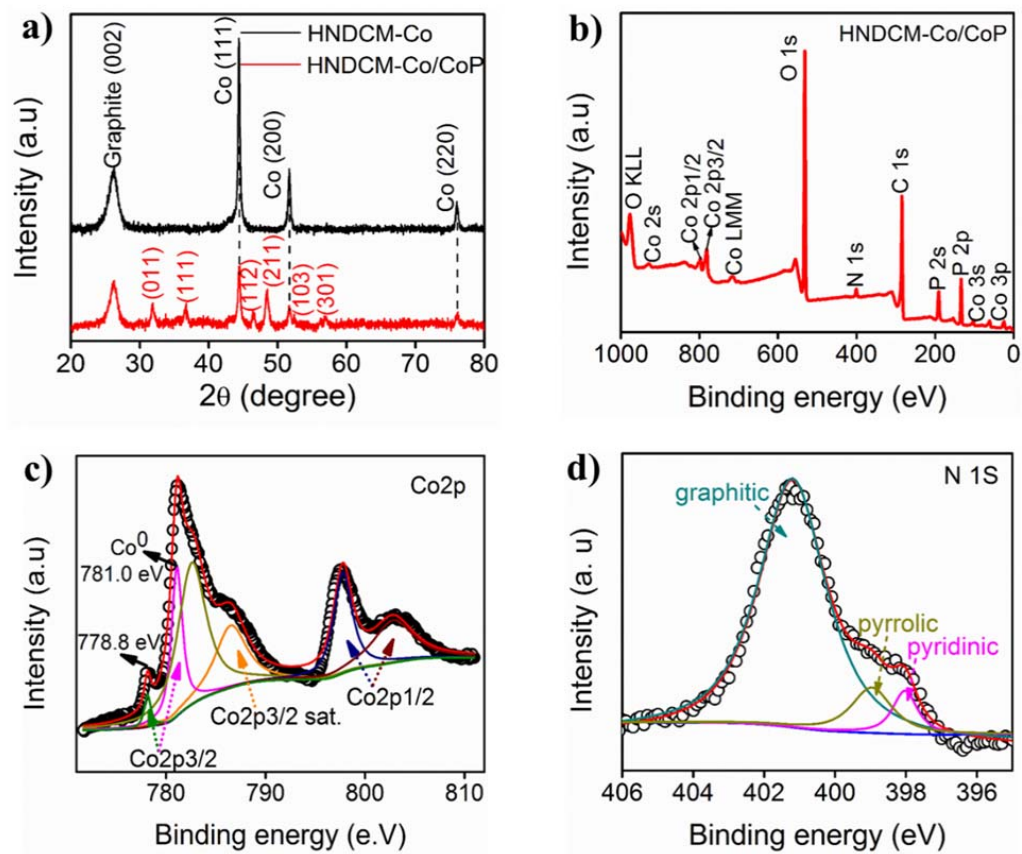

**Figure S5.** (a) XRD patterns of HNDCM-Co and HNDCM-Co/CoP; (b) XPS spectrum of HNDCM-Co/CoP; (c, d) XPS spectra of Co 2p and N 1s , respectively.

The phase structure of HNDCM-Co/CoP was further analyzed by X-ray diffraction (XRD) and X-ray photoelectron spectroscopy (XPS). Figure S6a shows the XRD patterns of HNDCM-Co and HNDCM-Co/CoP. The four peaks appearing in HNDCM-Co at 26º, 44º, 52º and 76º were indexed to the (002) reflection of graphite, metallic Co (111), Co (200) and Co (220), respectively. No other Co phase could be identified. After the phosphotization step to grow CoP, extra peaks at 32º, 36 º, 46 º, 48º, 52 º, and 58 º were found in HNDCM-Co/CoP, which are assigned to the (011), (111), (112), (211), (103), and (301) planes of CoP, respectively. Thus, the XRD data confirm that CoP is incorporated into the hybrid membrane. The XPS spectrum of HNDCM-Co/CoP (Figure S6b) further revealed the expected presence of Co, P, C, and N elements. The Co 2p high resolution XPS spectrum (Figure S5c) can be deconvoluted into two core-level signals, which are located at 780 and 796 eV, corresponding to Co2p$_{3/2}$ and Co2p$_{1/2}$, respectively. The peak at 781.0 eV is characteristic of Co$^0$ (ref. 1), accounting for 31 % of all Co species. The peak at 778.8 eV is typically assigned to the binding energies of Co $2p_{3/2}$ in CoP nanocrystal[2]. The total Co content was 42.8 mg/g as detected by inductively coupled plasma-atomic emission spectra. The relative content of Co and CoP in Janus-type Co/CoP nanocrystals can be calculated as 13.3 mg/g and 29.5 mg/g, respectively. The analysis here confirms the metallic Co and CoP coexist in the membrane. The N 1s peak (Figure S6d) can be deconvoluted into three different bands, 398.1, 399.5, and 400.7 eV, corresponding to the pyridinic (5.2 %), pyrrolic (8.9%) and graphitic (85.9%) nitrogen, respectively.

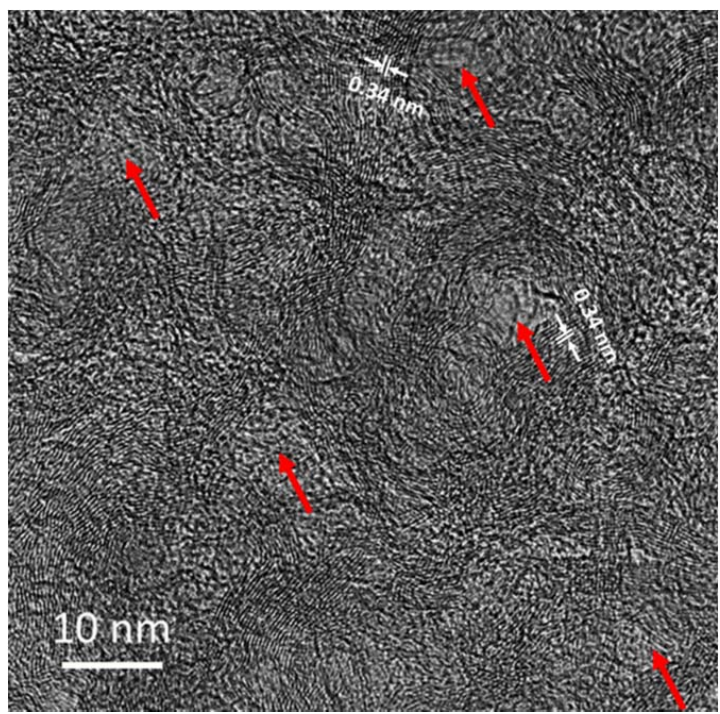

**Figure S6.** HRTEM image of the nitrogen-doped graphitic structure obtained from the HNDCM-Co/CoP sample. It can be clearly observed the presence of tiny mesopores (red line directed). The multishells of the pore wall are composed of graphitic layers with a d spacing of 0.34 nm, which is corresponding to the (002) plane of graphite.

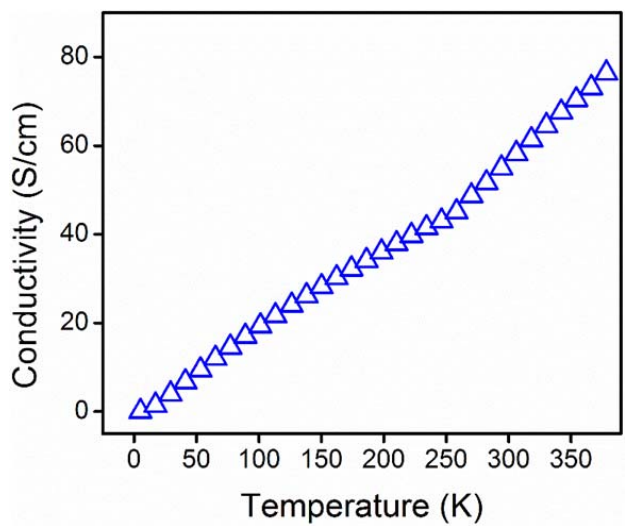

**Figure S7.** Temperature dependence of conductivity of HNDCM-Co/CoP measured by a four-probe method.

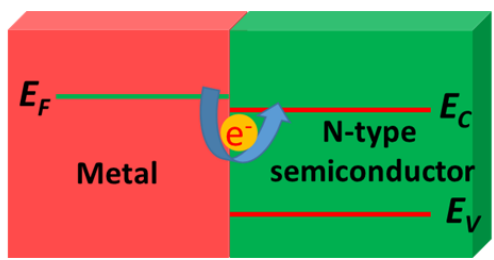

**Figure S8.** The electron transfer between metal and semiconductor in Janus nanoparticles *via* Mott–Schottky effect.

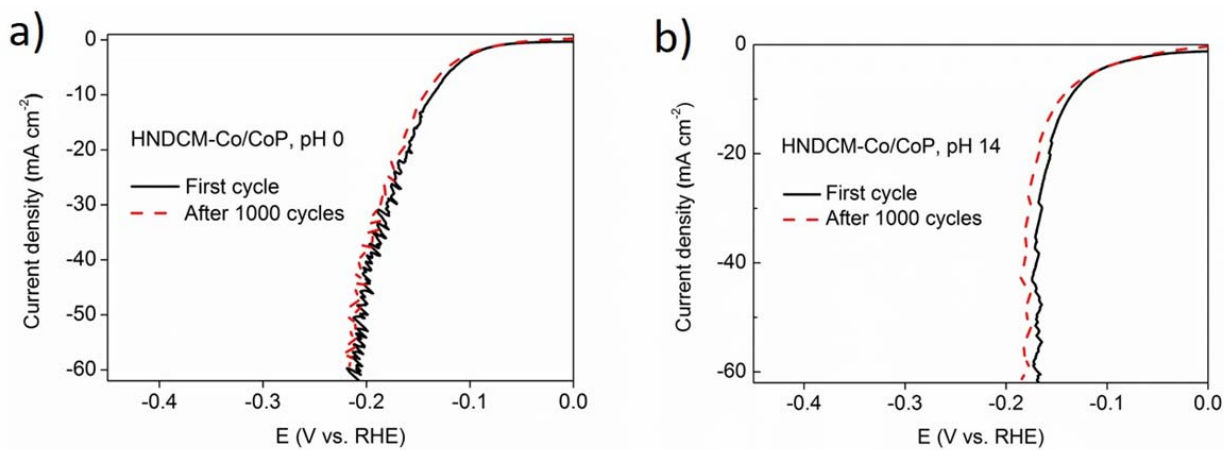

**Figure S9**. The accelerated cyclic voltammetry (CV) curves over 1000 cycles of HNDCM-Co/CoP in a) acid and b) alkaline conditions.

**Table S1**. HER performance of HNDCM-Co/CoP in this work, in comparison with several representative results with high performance non-noble metal based catalysts from recent publications.

| Catalyst | Current density j (mA cm$^{-2}$) | Overpotential (vs. RHE) at the corresponding j | Condition | *References* |
|---|---|---|---|---|
| MoB | 10 | 225 mV | Alkaline | *Angew. Chem., Int. Ed.* **2012**, *51*, 12703-12706. (S3) |
| MoC | 10 | > 250 mV | Alkaline | *Angew. Chem. Int. Ed.* **2014**, *126*, 6525-6528. (S4) |
| Co-NRCNT | 10 | 370 mV | Alkaline | *Angew. Chem., Int. Ed.* **2014**, *53*, 4461-4465. (S5) |
|  |  | 280 mV | Acid |  |
| CoO$_x$@CN | 10 | 232 mV | Alkaline | *J. Am. Chem. Soc.* **2015**, *137*, 2688-2694. (S1) |
| Nanoporous MoS$_2$ | 10 | 270 mV | Acid | *Nature Mater.* **2012**, *11*, 963-969. (S6) |
| Au supported MoS$_2$ | 0.2 | 150 mV | Acid | *Science* **2007**, *317*, 100-102. (S7) |
| Co−C−N Complex |  | 138 mV | Acid | *J. Am. Chem. Soc.* **2015**, *137*, 15070-15073. (S8) |
| Exfoliated WS$_2$/MoS$_2$ nanosheets | 10 | 187-210 mV | Acid | *Nature Mater.* **2013**, *12*, 850-855. (S9); *J. Am. Chem. Soc.* **2013**, *135*, 10274-10277. (S10) |
| MoS$_2$/Graphene | 10 | 150 mV | Acid | *J. Am. Chem. Soc.* **2011**, *133*, 7296-7299. (S11) |
| Oxygen-incorporated MoS$_2$ nanosheets | 10 | 180 mV | Acid | *J. Am. Chem. Soc.* **2013**, *135*, 17881-17888. (S12) |
| Co$_{0.6}$Mo$_{1.4}$N$_2$ | 10 | 200 mV | Acid | *J. Am. Chem. Soc.* **2013**, *135*, 19186-19192. (S13) |
| MoP | 10 | ~150 mV | Acid | *Energy Environ. Sci.* **2014**, *7*, 2624-2629. (S14) |
| CoSe$_2$ Nanoparticles/Carbon fiber paper | 10 | 137 mV | Acid | *J. Am. Chem. Soc.* **2014**, *136*, 4897-4900. (S15) |
| Ni$_{43}$Au$_{57}$ nanoparticles/carbon | 10 | ~200 mV | Acid | *J. Am. Chem. Soc.* **2015**, *137*, 5859-5862. (S16) |
| MnNi | 10 | 360 mV | Alkaline | *Adv. Funct. Mater.* **2015**, *25*, 393-399. (S17) |
| Co$_9$S$_8$@MoS$_2$ | 10 | 190 | acid | *Adv. Mater.* **2015**, *27*, 4752-4759 (S18) |
| CoS$_2$/RGO-CNT | 10 | 142 | acid | *Angew. Chem. Int. Ed.* **2014**, 53, 12594-12599 (S19) |
| CoS$_2$ | 10 | 145 | acid | *J. Am. Chem. Soc.* **2014**, 136, 10053-10061 (S20) |
| CoP nanowire array | 10 | 106 | Acid | *J. Am. Chem. Soc.* **2014**, 136, 7587-7590 (S21) |
| | | 209 | alkaline | |
| MoCN | 10 | 140 | acid | *J. Am. Chem. Soc.* **2015**, *137*, 110-113 (S22) |
| Ni$_2$P nanoparticles | 10 | 130 | acid | *J. Am. Chem. Soc.* **2013**, *135*, 9267-9270 (23) |
| Mn$_{0.03}$Co$_{0.97}$Se$_2$ | 10 | 218 | Acid | *J. Am. Chem. Soc.* **2016**, **138**, 5087−5092 (S24) |
| **HNDCM-Co/CoP** | **10** | **138 mV** | **alkaline** | ***in this work*** |
| | | **135 mV** | **acid** | |